\documentstyle[twoside,fleqn,espcrc2]{article}

 \renewcommand{\L}{L_b}
 \def\bbbone{{\mathchoice {\rm 1\mskip-4mu l} {\rm 1\mskip-4mu l}
 {\rm 1\mskip-4.5mu l} {\rm 1\mskip-5mu l}}}
 \newcommand{\subs}[1]{\mbox{$\!$\scriptsize\it#1}}
 \newcommand{\A}{{\cal A}}
 \newcommand{\Cstar}{C^{\ast}}
 \newcommand{\Dirac}{\mbox{$\not\!\!D$}}
 \newcommand{\Dm}{\triangle m^2}
 \newcommand{\G}{{\cal G}}
 \newcommand{\LAPN}{\Delta_{N,x}}
 \newcommand{\Lnull}{\Lambda^0}
 \newcommand{\Lone}{\Lambda^1}
 \newcommand{\Nc}{N_{\subs{c}}}

 \newcommand{\chin}{\chi^{(n)}}
 \newcommand{\equ}[1]{(\ref{#1})}
 \newcommand{\esk}{\enspace ,}
 \newcommand{\esp}{\enspace .}
 \newcommand{\gsim}{\raisebox{-3pt}{$\stackrel{>}{\sim}$}}
 \newcommand{\lambdanull}{\lambda_{0}(x)}
 \newcommand{\mcr}{m_{\subs{cr}}^2}
 \newcommand{\txeh}{{\textstyle\frac{1}{2}}}

\newcommand{\AmS}{{\protect\the\textfont2
  A\kern-.1667em\lower.5ex\hbox{M}\kern-.125emS}}

\title{Multigrid for propagators of staggered fermions in
       four-dimensional $SU(2)$ gauge fields}

\author{T. Kalkreuter\address{II. Institut f\"ur Theoretische Physik
        der Universit\"at Hamburg \\
        Luruper Chaussee 149, W-2000 Hamburg 50, Germany}%
        \thanks{Work supported by Deutsche Forschungsgemeinschaft.}}

\begin{document}

\begin{abstract}
  Multigrid (MG) methods for the computation of propagators of staggered
  fermions in non-Abelian gauge fields are discussed.
  MG could work in principle in arbitrarily disordered systems.
  The practical variational MG methods tested so far with a ``Laplacian
  choice'' for the restriction operator are not competitive with
  the conjugate gradient algorithm on lattices up to $18^4$.
  Numerical results are presented for propagators in $SU(2)$ gauge
  fields.
\end{abstract}

\maketitle

 \setcounter{footnote}{0}
\section{INTRODUCTION}

 In Monte Carlo simulations of lattice gauge theories with fermions
 the most time-consuming part is the computation of the gauge field
 dependent fermion propagators.
 Great hopes to compute propagators without any critical slowing down
 (CSD) are attached to multigrid (MG) methods
 \cite{BenBraSol,BRVGSP,BroRebVic,HSVLAT90,BERV,KalMacSpe,Vin,KalPL,%
       KalIJMPC,LauWit}.
 The MG methodology was reviewed by Brandt last year \cite{BraLAT91}.

 We wish to solve an equation
 \begin{equation}
   ( -\Dirac^2 + m^2 )\,\chi = f
 \end{equation}
 by MG, where $\Dirac$ is the gauge covariant staggered Dirac operator,
 and $m$ is a small quark mass.

 The following MG notations will be used.
 The fundamental lattice of lattice spacing $a_0$ is denoted by
 $\Lnull$.
 The first block lattice $\Lone$ has lattice spacing $a_1 = \L a_0$.
 Restriction and interpolation operators $C$ and $\A$\/, respectively,
 are given by kernels $C(x,z)$ and $\A (z,x)$ with $z \in \Lnull$,
 $x \in \Lone$\@.
 If $z$ is a site in block $x$, we write $z \in x$.
 Advantage is taken of the fact that we work in $d=4$ dimensions and
 that $\L =3$ will be chosen, so that only a two-grid algorithm was
 implemented.
 The residual equation on the coarse grid was solved exactly by
 the conjugate gradient (CG) algorithm.

\section{BLOCKED STAGGERED FERMIONS}

 We use a blocking procedure for staggered fermions which is
 consistent with the lattice symmetries of free fermions
 \cite{KalMacSpe}.
 This forces us to choose $\L =3$.
 Even $\L$ are not allowed.

 Fig.~\ref{FigBlockLattice} illustrates our choice of blocks.
 The different fermionic degrees of freedom are called ``pseudoflavor''
 \cite{KalMacSpe}.
 Different pseudoflavors are distinguished by different symbols in
 Fig.~\ref{FigBlockLattice}.
 Block centers $\hat{x}$ are encircled.
 The boundaries of seven blocks are marked.
 The averaging kernel $C(x,z)$ is only nonvanishing if site $z$ has the
 same pseudoflavor (symbol) as the block center $\hat{x}$.
 Therefore the seemingly overlapping blocks have actually no sites in
 common.
 \setlength{\unitlength}{0.60mm}
 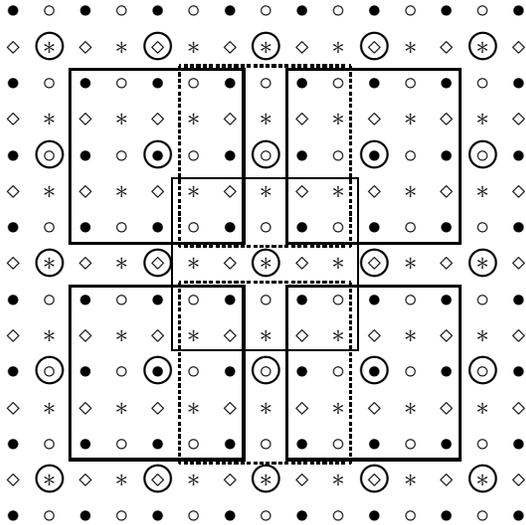
\begin{figure}[htb]
 \begin{minipage}{7.5 cm}{}
 \begin{center}
 \begin{picture}(115,115)(10.0,10.0)
 \multiput(10,10)(16,0){8}{$\bullet$}
 \multiput(18,10)(16,0){7}{$\circ$}
 \multiput(10,18)(16,0){8}{$\diamond$}
 \multiput(18,18)(16,0){7}{$\ast$}
 \multiput(10,26)(16,0){8}{$\bullet$}
 \multiput(18,26)(16,0){7}{$\circ$}
 \multiput(10,34)(16,0){8}{$\diamond$}
 \multiput(18,34)(16,0){7}{$\ast$}
 \multiput(10,42)(16,0){8}{$\bullet$}
 \multiput(18,42)(16,0){7}{$\circ$}
 \multiput(10,50)(16,0){8}{$\diamond$}
 \multiput(18,50)(16,0){7}{$\ast$}
 \multiput(10,58)(16,0){8}{$\bullet$}
 \multiput(18,58)(16,0){7}{$\circ$}
 \multiput(10,66)(16,0){8}{$\diamond$}
 \multiput(18,66)(16,0){7}{$\ast$}
 \multiput(10,74)(16,0){8}{$\bullet$}
 \multiput(18,74)(16,0){7}{$\circ$}
 \multiput(10,82)(16,0){8}{$\diamond$}
 \multiput(18,82)(16,0){7}{$\ast$}
 \multiput(10,90)(16,0){8}{$\bullet$}
 \multiput(18,90)(16,0){7}{$\circ$}
 \multiput(10,98)(16,0){8}{$\diamond$}
 \multiput(18,98)(16,0){7}{$\ast$}
 \multiput(10,106)(16,0){8}{$\bullet$}
 \multiput(18,106)(16,0){7}{$\circ$}
 \multiput(10,114)(16,0){8}{$\diamond$}
 \multiput(18,114)(16,0){7}{$\ast$}
 \multiput(10,122)(16,0){8}{$\bullet$}
 \multiput(18,122)(16,0){7}{$\circ$}
 \thicklines
 \multiput(19.5,19.7)(24,0){5}{\circle{6}}
 \multiput(19.5,43.7)(24,0){5}{\circle{6}}
 \multiput(19.5,67.7)(24,0){5}{\circle{6}}
 \multiput(19.5,91.7)(24,0){5}{\circle{6}}
 \multiput(19.5,115.7)(24,0){5}{\circle{6}}
 \put(24.25,24.25){\framebox(38,38){\mbox}}
 \put(24.25,72.25){\framebox(38,38){\mbox}}
 \put(72.25,24.25){\framebox(38,38){\mbox}}
 \put(72.25,72.25){\framebox(38,38){\mbox}}
 \put(48.25,23.25){\dashbox{0.7}(38,40){\mbox}}
 \put(48.25,71.25){\dashbox{0.7}(38,40){\mbox}}
 \thinlines
 \put(46.75,48.25){\framebox(41,38){\mbox}}
 \end{picture}
 \end{center}
 \end{minipage}
 \caption{Blocking of a two-dimensional staggered lattice consistent
          with the symmetries.}
 \label{FigBlockLattice}
 \end{figure}

\section{GROUND-STATE PROJECTION MG}

 The averaging kernel $C$ is chosen according to the ground-state
 projection (GSP) definition.
 The idea behind this definition is that the lowest mode of the
 fermion matrix, which is responsible for CSD in traditional algorithms,
 should be represented as well as possible on the block lattice.
 GSP was introduced in Refs. \cite{MacUnpublished,HSVGSP,BRVGSP}.
 An alternative MG approach to the inversion of the fermion matrix is
 the ``parallel-transported MG'' of Ben-Av et
 al.\ \cite{BenBraSol,LauWit}.

 In GSP the kernel $\Cstar$ of the adjoint of $C$ fulfills a gauge
 covariant eigenvalue equation,%
 \footnote{$C$ depends on the gauge field although this is not
           indicated explicitly.}
 \begin{equation}
   ( -\LAPN \Cstar ) (z,x) = \lambdanull\,\Cstar (z,x) \esp
 \label{EVequationC}
 \end{equation}
 $\lambdanull$ is the lowest (gauge invariant) eigenvalue of the
 positive (semi-)definite operator $-\LAPN$ which will be specified
 below.
 $\Cstar (z,x) = C(x,z)^{\dagger}$, where $^{\dagger}$ denotes
 Hermitean conjugation, as usual.
 The solution of Eq.~\equ{EVequationC} is made unique by imposing
 the normalization condition
 \begin{equation}
   C \Cstar = \bbbone
 \label{Normalization}
 \end{equation}
 and the covariance condition
 \begin{equation}
   C(x,\hat{x}) = r(x)\,\bbbone
 \label{CovarianceCondition}
 \end{equation}
 with $r(x) > 0$.
 (In case of gauge groups $\G$ different from $U(1)$ or $SU(2)$ the
  r.h.s.\ of \equ{CovarianceCondition} is replaced by a positive
  Hermitean matrix.)
 The covariance condition \equ{CovarianceCondition} ensures that
 the kernel $C(x,z)$ transforms covariantly under gauge transformations.
 In the limiting case of vanishing gauge coupling, GSP yields
 (gauge-transformed) piecewise constant kernels on blocks.

 In a gauge theory with $\Nc$ colors, $C(x,z)$ is an $\Nc\times\Nc$
 matrix which will not be an element of $\G$ in general.
 For $\G = SU(2)$ it is easy to see that 1-column
 vector solutions of \equ{EVequationC} may be combined into 2-column
 matrix solutions \cite{KalPL}.

\subsection{Bosons}

 In case of bosons $\LAPN$ is the covariant Laplacian with Neumann
 boundary conditions on boundaries of blocks $x$.
 $\LAPN$ is defined by
 \begin{eqnarray}
  & & (\LAPN \Cstar ) (z,x) = \nonumber\\
  & &             \sum_{\stackrel{{\scriptstyle z'\ \subs{\rm n.n.}\ z}}
                                 {z' \in x}}
                 \left[ U(z,z') \Cstar (z',x) - \Cstar (z,x) \right]
 \label{LAPN}
 \end{eqnarray}
 for $z \in x$.
 $U(z,z')$ is the gauge field on the link $(z,z')$.
 As a supplement we enforce Dirichlet boundary conditions in such a way
 that $C(x,z) = 0$ unless $z \in x$.

\subsection{Staggered fermions}

 For fermions $\Delta$ in \equ{EVequationC} was chosen as the gauge
 covariant fermionic ``2-link Laplacian'' which is defined through
 $-\Dirac^2 = -\Delta + \sigma_{\mu\nu} F_{\mu\nu}$ \cite{KalMacSpe}.
 With this ``Laplacian choice'' $C(x,z)$ is only nonvanishing when
 $z$ and $x$ carry the same pseudoflavor.
 Other proposals were made in Ref.~\cite{KalMacSpe}, but have not
 been implemented yet.

 There exists an efficient algorithm for the solution of
 Eq.~\equ{EVequationC} \cite{KalNP2}.
 Therefore it is not necessary to sacrifice the gauge covariance of GSP
 by wasting computer time for gauge fixing.

\subsection{Idealized MG algorithm}

 Given the averaging kernel $C$ there exists an associated idealized
 interpolation kernel $\A$\/.
 By means of this $\A$ a (nearly) critical system on $\Lnull$ is
 mapped onto a noncritical system on the MG\@.
 The construction of the optimal $\A$ was described in
 Mack's Carg\`{e}se lectures \cite{MacCargese}.
 Its origin are works on constructive quantum field theory.

 The gauge covariant optimal $\A$ is the solution of the equation
 \begin{equation}
  \left( [ D + \kappa\,\Cstar C ] \A \right) (z,x)
  = \kappa\,\Cstar (z,x)
 \label{optimalA}
 \end{equation}
 for large $\kappa$.
 $D = -\Delta + m^2$ for bosons, and $D= -\Dirac^2 + m^2$ for fermions.

 An averaging kernel $C$ is said to be ``good'' (or to define a ``good''
 block spin) if the associated $\A (z,x)$ decays exponentially in $| z
 - \hat{x} |$.
 GSP defines a good $C$ in arbitrarily disordered gauge fields, both
 for bosons and for staggered fermions.
 Fig.~\ref{FigAkernel} visualizes an example.
 \begin{figure}[htb]
 \vspace*{5.6cm}
 \caption{An optimal fermionic interpolation kernel $\A (z,0)$ in a
          quenched $SU(2)$ gauge field at $\beta=2.7$ on an $18^4$
          lattice.
          Shown is a two-dimensional cut through the block center $x=0$;
          $z_3$ and $z_4$ are fixed.
          The vertical axis gives the trace norm of $\A (z,x)$.}
 \label{FigAkernel}
 \end{figure}
 $\A$ decays exponentially over distance $L_b a_0$ (nearly) as fast
 in the presence of gauge fields than in their absence.

 By means of the idealized $\A$ it was shown in Ref.~\cite{KalPL}
 that MG computations of bosonic propagators without CSD are possible
 in arbitrarily disordered gauge fields
 (including the case $\beta = 0$).
 These computations showed for the first time that the MG method could
 work in principle in arbitrarily strong disorder.
 Unfortunately, the idealized MG algorithm is not practical for
 production runs because of computational complexity and storage
 space requirements, but it was important to answer questions of
 principle.

\subsection{Variational MG}

 A practical MG method is variational coarsening.
 There $\A = \Cstar$, and the coarse grid operator is $C D \Cstar$.
 This method reduces to piecewise constant interpolation in the
 absence of gauge fields, where it is known that CSD is eliminated.

 However, in nontrivial gauge fields variational MG with the Laplacian
 choice of $C$ is not competitive with CG on lattices up to $18^4$;
 see Fig.~\ref{FigMGCGcomparison}.
 \begin{figure}[htb]
 \vspace*{4.2cm}
 \caption{Computation of propagators of staggered fermions
         $(-\Dirac^2 + m^2 )^{-1}$ on an $18^4$ lattice in a quenched
         $SU(2)$ gauge field at $\beta = 2.7$.
         $r$~denotes the residual.
         The numbers refer to the following algorithms:
         1/2: variational MG SOR (lexicographic ordering) with $m^2 =
              0.01$/$0.0$, $\omega = 1.88$/$1.92$;
         3/4: CG with $m^2 = 0.01$/$0.0$.}
 \label{FigMGCGcomparison}
 \end{figure}
 The non-optimized MG program requires a factor of $4.5$ more
 arithmetic operations than CG\@.

 A further numerical finding is the following.
 Relaxation times $\tau$ of conventional one-grid relaxation and of
 variational MG with the Laplacian choice of $C$ follow a scaling
 relation%
 \footnote{In case of one-grid relaxation for bosons
           Eq.~\equ{ScalingTau} is known analytically.}
 \begin{equation}
   \tau = \frac{\mbox{{\em const.}}}{\Dm} \quad \mbox{with} \quad
   \Dm = m^2 - \mcr \esk
 \label{ScalingTau}
 \end{equation}
 where $\mcr$ is the lowest eigenvalue of $-\Dirac^2$, and {\em const.}
 is {\em independent of the lattice size\/}.
 $\mcr$ is small and is usually neglected, but on relatively small
 lattices this neglect is not justified.
 From \equ{ScalingTau} we conclude that variational MG with the
 Laplacian choice of $C$ will not be able to eliminate CSD on large
 lattices.

\section{UPDATING ON AN MG LAYER CON\-SISTING OF A SINGLE SITE}

 Because of \equ{ScalingTau}, (MG) relaxation continues to have CSD even
 on a lattice of only $2^d$ sites.
 Therefore it seems necessary to update on a ``last site'' in order to
 eliminate the appearance of CSD\@.

 Given an approximate propagator $\chin$, this updating (in the
 unigrid point of view) amounts to globally rescaling $\chin$ by a gauge
 covariant $\Nc \times \Nc$ matrix $\Omega$ (in case of bosons or
 Wilson fermions) \cite{KalIJMPC,Kalthesis}:
 \begin{equation}
    \chin (z) \mapsto \chin (z)\,\Omega \esp
 \label{rescaleBOSE}
 \end{equation}
 The matrix $\Omega$ is chosen such that the energy functional
  $K [ \chi ] = \txeh < \chi\,,\,D\,\chi >\ - < \chi\,,\,f >$
 of the rescaled approximation $\chin\,\Omega$ is minimized.

 In case of staggered fermions we have to consider that there are $2^d$
 different pseudoflavors, and we replace $\Omega$ in \equ{rescaleBOSE}
 by $\Omega (H(z))$ where $H(z)$ denotes the pseudoflavor of $z$.
 The $2^d$ $\Nc\times\Nc$ matrices $\Omega (H)$ are again chosen
 automatically by the algorithm in such a way that $K [ \chin \Omega ]$
 gets minimized.

 We note that $\Omega$ and the $\Omega (H)$'s practically equal
 $\bbbone$ as soon as errors decay exponentially.

\subsection{Results for bosons}

 In a fixed volume the asymptotic fall-off properties do not depend
 on $\Dm$ when \equ{rescaleBOSE} is included in algorithms.
 When the lattice size is increased, the asymptotic fall-off rate
 does not change, there is only a mild volume effect
 with respect to how long it takes until errors decay exponentially.
 This volume effect is much milder than the one in (unpreconditioned)
 CG, and variational MG plus \equ{rescaleBOSE} begins to become
 superior in computer time for lattices $\gsim 18^4$.

\subsection{Results for staggered fermions}

 In a fixed volume the {\em asymptotic} $1 / \Dm$ divergence is
 eliminated when the analog of \equ{rescaleBOSE} is included in
 algorithms.
 However, it takes a relatively large number of iterations until
 fall-offs are exponential.
 Also, in contrast to the bosonic case, asymptotic decay rates
 increase at fixed $\Dm$ when the lattice size is increased.
 Therefore CSD seems not to be eliminated, and it is questionable
 whether the method pays in practice on lattices of realizable sizes.
 For details see Refs. \cite{KalIJMPC,Kalthesis}.

\section{CONCLUSIONS}

 The variational MG methods tested so far for staggered fermions
 are not competitive with CG.
 We expect better results when blocks are used which overlap in
 a nontrivial way.
 The question, whether it is sufficient to average only over sites
 with the same pseudoflavor in nontrivial gauge fields, also has to be
 investigated.
 The Diracian proposal for $C$ of Ref.~\cite{KalMacSpe} should be
 tested.

\smallskip\noindent{ACKNOWLEDGMENTS}

 I am indebted to G.\,Mack for many stimulating discussions.
 Financial support by Deutsche Forschungsgemeinschaft
 is gratefully acknowledged.
 For providing resources and help I wish to thank
 HLRZ J\"ulich and its staff.

\end{document}